\documentclass[letter,longauth]{aa}  
\usepackage{graphicx}	% Including figure files
\usepackage{multirow}
\usepackage{txfonts}
\usepackage[colorlinks, citecolor=blue, linkcolor=blue]{hyperref}
\newcommand{\nhh}{$n_{\mathrm{H_2}}$}

\begin{document} 

   \title{A constant N$_2$H$^+$\,(1-0)-to-HCN\,(1-0) ratio on kiloparsec scales}

   \author{M.~J.~Jim\'enez-Donaire
          \inst{\ref{oan},\ref{yebes}} \fnmsep\thanks{\email{mdonaire@oan.es}}
          \and 
          A.~Usero \inst{\ref{oan}}         
          \and
          I.~Be{\v{s}}li{\'c} \inst{\ref{paris}}
          \and
          M.~Tafalla \inst{\ref{oan}}
          \and
          A.~Chac\'on-Tanarro \inst{\ref{oan}}
          \and
          Q.~Salom\'e \inst{\ref{turku},\ref{metsahovi}}
          \and
          C.~Eibensteiner \inst{\ref{bonn}}
          \and
          A.~Garc\'ia-Rodr\'iguez \inst{\ref{oan}}
          \and
          A.~Hacar \inst{\ref{vienna}}
          \and 
          A.~T.~Barnes \inst{\ref{eso}}     
          \and 
          F.~Bigiel \inst{\ref{bonn}}          
          \and 
          M.~Chevance \inst{\ref{ita}}
          \and
          D.~Colombo \inst{\ref{bonn}}
          \and 
          D.~A.~Dale \inst{\ref{wyoming}}
          \and
          T.~A.~Davis \inst{\ref{cardiff}}
          \and
          S.~C.~O.~Glover \inst{\ref{ita}}
          \and 
          J.~Kauffmann \inst{\ref{mit}}          
          \and 
          R.~S.~Klessen \inst{\ref{ita}, \ref{iwr}}
          \and 
          A.~K.~Leroy \inst{\ref{osu}}         
          \and
          L.~Neumann \inst{\ref{bonn}}
          \and 
          H.~Pan \inst{\ref{taipei}}
          \and
          J.~Pety \inst{\ref{iramfr},\ref{paris}}
          \and
          M.~Querejeta \inst{\ref{oan}}
          \and
          T.~Saito \inst{\ref{naoj}}      
          \and
          E.~Schinnerer \inst{\ref{mpia}}
          \and
          S.~Stuber \inst{\ref{mpia}}
          \and
          T.~G.~Williams \inst{\ref{oxford}}
          }

   \institute{Observatorio Astronómico Nacional (IGN), C/Alfonso XII, 3, E-28014 Madrid, Spain \label{oan}\\
              \email{mdonaire@oan.es}
         \and
             Centro de Desarrollos Tecnológicos, Observatorio de Yebes (IGN), 19141 Yebes, Guadalajara, Spain \label{yebes}
        \and
             LERMA, Observatoire de Paris, PSL Research University, CNRS, Sorbonne Universit\'es, 75014 Paris, France \label{paris}
        \and
            Finnish Centre for Astronomy with ESO (FINCA), University of Turku, Vesilinnantie 5, 20014 Turku, Finland \label{turku}
        \and
            Aalto University Mets\"ahovi Radio Observatory, Mets\"ahovintie 114, 02540 Kylm\"al\"a, Finland \label{metsahovi}
        \and   
            Argelander-Institut f\"ur Astronomie, Universit\"at Bonn, Auf dem H\"ugel 71, 53121 Bonn, Germany \label{bonn}            
        \and
            Department of Astrophysics, University of Vienna, T\"urkenschanzstrasse 17, A-1180 Vienna, Austria \label{vienna}
        \and   
            European Southern Observatory (ESO), Karl-Schwarzschild-Stra{\ss}e 2, 85748 Garching, Germany \label{eso}            
        \and
            Universit\"{a}t Heidelberg, Zentrum f\"{u}r Astronomie, Institut  f\"{u}r Theoretische Astrophysik, Albert-Ueberle-Strasse 2, 69120 Heidelberg, Germany \label{ita}
        \and
            Department of Physics \& Astronomy, University of Wyoming, Laramie, WY USA \label{wyoming}
         \and
            Cardiff Hub for Astrophysics Research \&\ Technology, School of Physics \&\ Astronomy, Cardiff University, Queens Buildings, Cardiff, CF24 3AA, UK \label{cardiff}
        \and
            Haystack Observatory, Massachusetts Institute of Technology, 99 Millstone Rd., Westford, MA 01886, USA \label{mit}            
         \and
            Universit\"{a}t Heidelberg, Interdisziplin\"ares  Zentrum f\"{u}r Wissenschaftliches Rechnen, Im Neuenheimer Feld 225, 69120 Heidelberg, Germany \label{iwr}
        \and
            Department of Astronomy, The Ohio State University, 140 West 18th Ave, Columbus, OH 43210, USA \label{osu}
        \and
            Department of Physics, Tamkang University, No.151, Yingzhuan Road, Tamsui District, New Taipei City 251301, Taiwan \label{taipei}
        \and 
            Institut de Radioastronomie Millim\'etrique (IRAM), 300 Rue de la Piscine, F-38406 Saint Martin d’Hères, France \label{iramfr}
         \and             
            National Astronomical Observatory of Japan, 2-21-1 Osawa, Mitaka, Tokyo, 181-8588, Japan \label{naoj}
         \and
            Max Planck Institute for Astronomy, K\"onigstuhl 17, D-69117 Heidelberg, Germany \label{mpia}
        \and
            Sub-department of Astrophysics, Department of Physics, University of Oxford, Keble Road, Oxford OX1 3RH, UK \label{oxford}
            \\
             }
             
   \date{Received 30 May 2023; accepted 31 July 2023}

% \abstract{}{}{}{}{} 
% 5 {} token are mandatory
 
  \abstract{Nitrogen hydrides such as NH$_3$ and N$_2$H$^+$ are widely used by Galactic observers to trace the cold dense regions of the interstellar medium. In external galaxies, because of limited sensitivity, HCN has become the most common tracer of dense gas over large parts of galaxies. We provide the first systematic measurements of N$_2$H$^+$\,(1-0) across different environments of an external spiral galaxy, NGC\,6946. We find a strong correlation ($r>0.98\,,p<0.01$) between the HCN\,(1-0) and N$_2$H$^+$\,(1-0) intensities across the inner $\sim8\,\mathrm{kpc}$ of the galaxy, at kiloparsec scales. This correlation is equally strong between the ratios N$_2$H$^+$\,(1-0)/CO\,(1-0) and HCN\,(1-0)/CO\,(1-0), tracers of dense gas fractions ($f_\mathrm{dense}$). We measure an average intensity ratio of N$_2$H$^+$\,(1-0)/HCN\,(1-0)$\,=0.15\pm0.02$ over our set of five IRAM-30m pointings. These trends are further supported by existing measurements for Galactic and extragalactic sources. This narrow distribution in the average ratio suggests that the observed systematic trends found in kiloparsec-scale extragalactic studies of $f_\mathrm{dense}$ and the efficiency of dense gas ($\mathrm{SFE}_\mathrm{dense}$) would not change if we employed N$_2$H$^+$\,(1-0) as a more direct tracer of dense gas. At kiloparsec scales our results indicate that the HCN\,(1-0) emission can be used to predict the expected N$_2$H$^+$\,(1-0) over those regions. Our results suggest that, even if HCN\,(1-0) and N$_2$H$^+$\,(1-0) trace different density regimes within molecular clouds, subcloud differences average out at kiloparsec scales, yielding the two tracers proportional to each other.}


   \maketitle

%-------------------------------------------------------------------

\section{Introduction}

Understanding the link between star formation and the available molecular reservoir is crucial to comprehending how gas density regulates star formation over cosmic time. While a variety of molecular lines are commonly employed to infer properties of the sites of current or future star formation, the rotational transitions of carbon monoxide (CO) have been the preferred tracers of the bulk molecular ISM in the Milky Way and other galaxies \citep[e.g.,][]{Dame1987,Bally1987,Kuno2007} because of the CO abundance and low critical density ($\sim10^2\,\mathrm{cm}^{-3}$).

Star formation has been shown to take place in the densest parts of molecular clouds \citep[e.g.,][]{Heiderman2010,Lada2010,Lada2012,Evans2014}. These dense and compact regions are difficult to resolve in external galaxies. Lines that are more difficult to excite, such as HCN\,(1-0), and their ratio to low-density tracers, such as low-$J$\,CO, offer the best way to routinely probe the physical conditions and distribution of dense gas in external galaxies \citep[e.g.,][]{Leroy2017}. Over the last decade, multiple extragalactic surveys have mapped the dense gas content at low resolution  \citep[kiloparsec scales; e.g., EMPIRE][]{Bigiel2016,Donaire2019} and high resolution \citep[parsec scales; e.g.,][]{Gallagher2018,Querejeta2019,Beslic2021,SanchezGarcia22,Eibensteiner2022}. All these studies share a key result: the HCN-to-IR ratio, an observational proxy for the star formation efficiency of dense gas (SFE$_\mathrm{dense}$), depends on the host galaxy and local environment. While these observed trends are in contrast to a hypothesized constant SFE$_\mathrm{dense}$ above some critical surface density as observed on much smaller scales in the Milky Way \citep[e.g.,][]{Lada2010,Lada2012,Evans2014}, this conclusion critically rests on our ability to estimate dense gas masses.

In addition to dust continuum emission, nitrogen hydrides, such as NH$_3$ and N$_2$H$^+$, are widely used in Galactic observations to trace cold and dense regions. N$_2$H$^+$ is a very selective high-density tracer, as it only efficiently forms once CO is heavily depleted \citep{Bergin2007}, and therefore traces cold dense gas. Recent studies in the Galaxy compared the emission of dense gas tracers including HCN and N$_2$H$^+$, with dust temperatures or visual extinction \citep[e.g.,][]{Kauffmann2017,Pety2017,Barnes2020,Evans2020,Patra2022,Dame2023ApJ}. The results in Orion, Perseus, and W49 \citep{Kauffmann2017,Pety2017,Tafalla2021} show that N$_2$H$^+$ is the only emission line among the dense gas surveys that remains undetected at low densities ($N_{\mathrm{H_2}}<10^{22}$\,cm$^{-2}$, \nhh<$10^4$\,cm$^{-3}$), but efficiently emits at higher densities, becoming the only truly density-selective tracer in their sample.

Because radio observations of NH$_3$ and N$_2$H$^+$ in different extragalactic environments are very costly and sensitivity limited, previous works have mainly focused on the brightest targets: active galactic nuclei (AGNs), starburst galaxies, and ultra-luminous infrared galaxies (ULIRGs) \citep[e.g.,][]{Mauersberger1991,Meier2005,Meier2012,Watanabe2014,Aladro2015,Martin2021}. We present the first extragalactic observations of N$_2$H$^+$\,(1-0) and HCN\,(1-0) in a wide range of dynamical conditions and star formation properties sampled across an entire galaxy disk, NGC\,6946. This nearby source \citep[$7.72$\,Mpc,][]{Anand2018Distance}  is a gas-rich, very actively star-forming double-barred spiral galaxy. The five positions observed cover very diverse environments within a galaxy disk. These include a central starburst, spiral and interarm regions, as well as the outer disk. The selected regions span 1.7\,dex (a factor of $\sim50$) in star formation rate (SFR) surface density. The observations presented here provide a crucial benchmark to understand how well HCN performs as a tracer of cold dense gas that favors N$_2$H$^+$ emission, in a representative external galaxy. 

\begin{figure}
    \centering
    \includegraphics[width=\columnwidth]{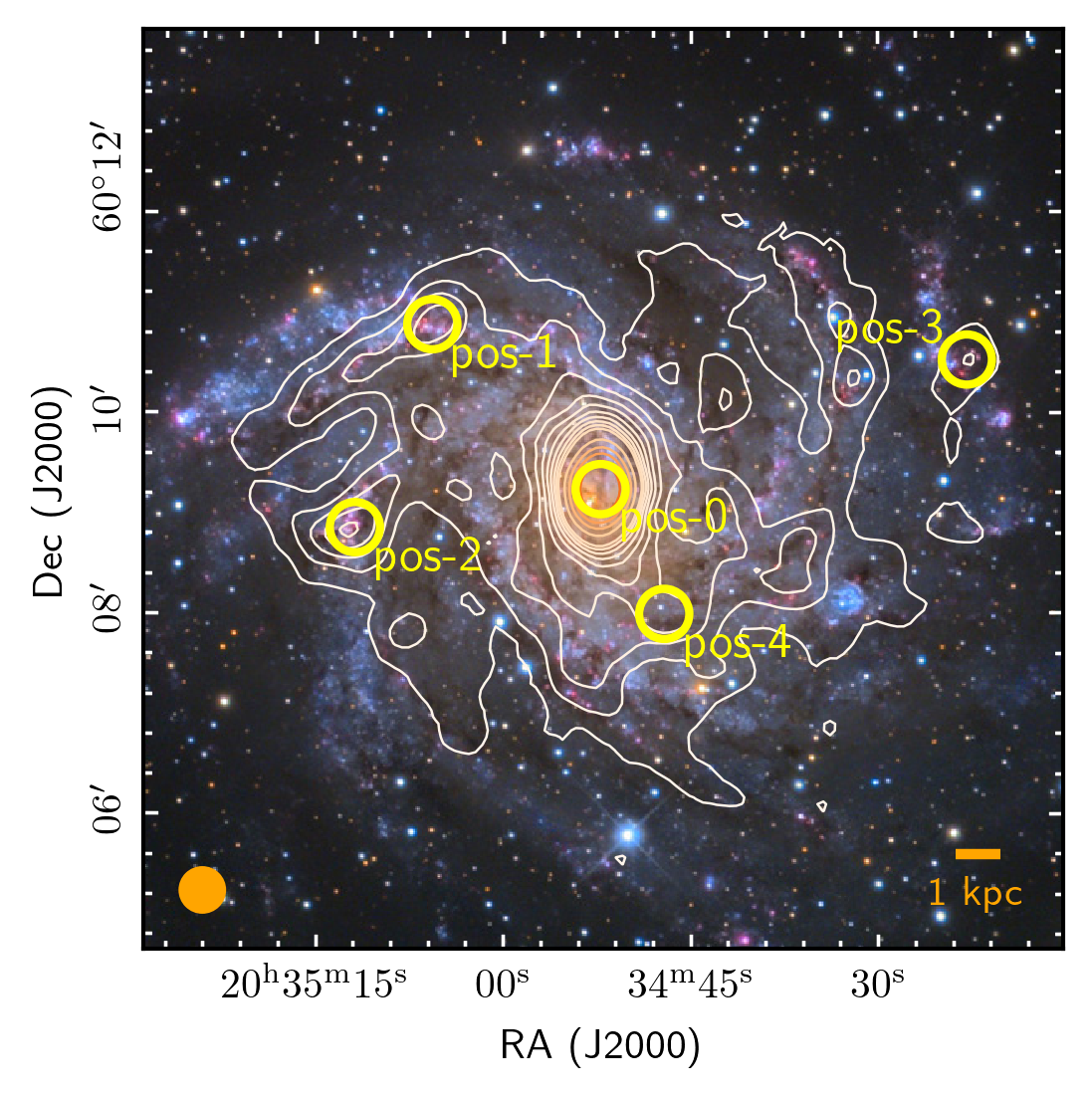}
    \caption{{\bf NGC\,6946 with EMPIRE CO\,(1-0) contours} overlaid on an optical image. The contours illustrate the IRAM-30m $^{12}$CO\,(1-0) integrated intensities. The millimeter observations have a resolution of $26$\arcsec (${\sim}800\,\rm pc$), indicated by the orange circle. The contours are drawn at arbitrary intervals between $5$ and $200$\,K\, km\,s$^{-1}$ to highlight the distribution of the molecular gas disk. The yellow circles indicate the locations of our pointed N$_2$H$^+$\,(1-0) and HCN\,(1-0) observations, covering a wide range of galactic environmental conditions: center (0), spiral arms (1 and 2), interarm (4), and outskirts (3). The size represents the  $28\arcsec$ resolution of the IRAM-30m beam at $\sim86\,$GHz.}
    \label{fig:composite}
\end{figure}

\begin{table}
    \begin{center}
    \caption{{Main properties of NGC\,6946}}
    \label{tab:properties}
    \begin{tabular}{l c c}
    \hline \hline
    Parameter            &  Value &  Notes \\ \hline
    R.A. (J2000)         & $20{:}34{:}52.35$  & (1)  \\
    Decl. (J2000)        & $+60{:}09{:}14.58$ &  (1) \\
    Morphology           & SAB(rs)cd         &   (2) \\
    Nuclear type         & star-forming, H\,II  & (3) \\
    Distance            & $7.72$ Mpc      &   (4) \\
    Inclination          & $33\degr$        &  (5) \\
    P.A. major axis      & $242\degr$        &   (6) \\
    V$_{\rm LSR}$        & $50$ km\,s$^{-1}$&  (1)  \\
    $\mathrm{SFR}$  & $6.17$\, M$_\odot$\,yr$^{-1}$ & (7)\\
    $\log_{10}(M_\star)$ & $10.39\,\log_{10}(\mathrm{M}_\odot)$ & (7)\\ \hline
    \hline \hline
    \end{tabular}
    \end{center}
    \begin{minipage}{0.95\columnwidth}
        \vspace{1mm}
        {\bf Notes:} (1): \cite{Schinnerer2006}; (2) \cite{deVaucouleurs1991}; (3): \cite{Goulding2009}; (4): \cite{Anand2018Distance}; (5): \cite{deBlok2008Inclination}; (6): \cite{Crosthwaite2002}; (7): \cite{Leroy2019}.
    \end{minipage}
\end{table}

%--------------------------------------------------------------------%--------------------------------------------------------------------
%--------------------------------------------------------------------%--------------------------------------------------------------------
\section{Observations and data reduction}

\begin{table*}[h!]
    \begin{center}
    \caption{Summary of observations and spectral line parameters for each observed position.}
    \label{tab:observations}
    \begin{tabular}{c c c c c c c c c c c c}
         \hline \hline 
          Position & RA (J2000) & DEC (J2000) & Radius & Line & Int. intensity & FWHM & $V_{lsr}$ & $\langle \rm rms\rangle$& Obs. time & $\langle T_{\rm sys}\rangle$\\
         & [hh:mm:ss] & [$^o$:$\arcmin$:$\arcsec$] & [kpc] & ($1-0$) & [K~km~s$^{-1}$] & [km~s$^{-1}$] & [km~s$^{-1}$] & [mK] & [hr] & [K]\\
         & & & & & & & &(1)&(2)&(3)\\\hline
         \multirow{2}{*}{Pos-0} & \multirow{2}{*}{20:34:52.20} & \multirow{2}{*}{+60:09:14} & \multirow{2}{*}{0.0} & HCN & $11.12\pm0.01$ & $147\pm15$ & $56\pm6$ & 0.3 & \multirow{2}{*}{44.3} & \multirow{2}{*}{92}\\
         &&&&N$_2$H$^+$ & $1.83\pm0.01$ & $133\pm11$ & $55\pm6$ & 0.2\\
      \hline   
         \multirow{2}{*}{Pos-1} & \multirow{2}{*}{20:35:05.70} & \multirow{2}{*}{+60:10:52} & \multirow{2}{*}{5.1} & HCN & $0.33\pm0.01$ & $32\pm2$ & $-32\pm4$ & 0.5 & \multirow{2}{*}{10.0} & \multirow{2}{*}{112}\\
         &&&&N$_2$H$^+$ & $0.040\pm0.008$ & $21\pm6$ & $-28\pm3$ & 0.4\\
\hline
         \multirow{2}{*}{Pos-2} & \multirow{2}{*}{20:35:11.88} & \multirow{2}{*}{+60:08:51} & \multirow{2}{*}{5.7} & HCN & $0.39\pm0.01$ & $31\pm2$ & $-9\pm2$ & 0.6 & \multirow{2}{*}{8.5} & \multirow{2}{*}{116}\\
         &&&&N$_2$H$^+$ & $0.067\pm0.011$ & $27\pm5$ & $-6\pm2$ & 0.5\\
\hline
         \multirow{2}{*}{Pos-3} & \multirow{2}{*}{20:34:22.86} & \multirow{2}{*}{+60:10:31} & \multirow{2}{*}{8.6} & HCN & $0.13\pm0.01$ & $25\pm2$ & $114\pm12$ & 0.5 & \multirow{2}{*}{13.9} & \multirow{2}{*}{108}\\
         &&&&N$_2$H$^+$ & $0.026\pm0.006$ & $30\pm8$ & $117\pm15$ & 0.3\\
\hline
         \multirow{2}{*}{Pos-4} & \multirow{2}{*}{20:34:47.10} & \multirow{2}{*}{+60:07:59} & \multirow{2}{*}{3.3} & HCN & $0.26\pm0.01$ & $32\pm4$ & $126\pm12$ & 0.4 & \multirow{2}{*}{13.2} & \multirow{2}{*}{99}\\
         &&&&N$_2$H$^+$ & $0.031\pm0.007$ & $20\pm6$ & $131\pm15$ & 0.3\\         
         
         \hline
    \end{tabular}
    \end{center}
    {\raggedright {\bf Notes:} (1) Average rms measured in a 4\,km\,s$^{-1}$ wide channel. (2) Total on-source time used to generate the final spectra after reduction. (3) Average system temperature during observations.}
\end{table*}

We observed NGC\,6946 with the IRAM-30m telescope from 2018 to 2020, using the 3\,mm band (E090) of the dual-polarization Eight MIxer Receiver \citep[EMIR,][]{Carter2012}. The observations presented here correspond to a total $\sim90\,\mathrm{h}$ of on-source observing time (see Table \ref{tab:observations}). We achieved an angular resolution of $\sim28\arcsec$ at 86\,GHz. Additional details on the observations, as well as N$_2$H$^+$\,(1-0) and HCN\,(1-0) data reduction and analysis are given in Appendix \ref{ap:datareduction}.

NGC\,6946 belongs to the IRAM-30m survey EMPIRE \citep{Donaire2019}, and thus a suite of high critical density tracers as well as CO\,(1-0) and carbon isotopologs are publicly available. These lines were mapped at $\sim 30\arcsec$ resolution across a total area of $4.5\,\arcmin\times6.5\,\arcmin$; therefore, a direct comparison to common dense and bulk molecular gas tracers is possible. For this study we used the CO\,(1-0) EMPIRE integrated intensity maps of NGC\,6946 \citep{Donaire2019} to compare the line ratios derived for each observed position in Fig.\,\ref{fig:composite}.

\section{Results}

N$_2$H$^+$\,(1-0) and HCN\,(1-0) were successfully detected at every position with a peak signal-to-noise ratio higher than or equal to 3. The spectra presented in Fig.\,\ref{fig:spectra} show that the detected line centroid of N$_2$H$^+$\,(1-0) agrees with the mean HCN\,(1-0) velocity, and the two transition lines show similar line widths within $\leq5\,\%$, indicating that the presence of HCN-emitting gas is always accompanied by gas emitting N$_2$H$^+$ in our $\sim 28\arcsec$ beam ($\sim 1\,$kpc). It is not possible to resolve the hyperfine structure (hfs) of these lines in the observed positions since the splitting separation ($\sim12\,\mathrm{km \, s}^{-1}$ for HCN\,(1-0) and $\sim15\,\mathrm{km \, s}^{-1}$ for N$_2$H$^+$\,(1-0)) is smaller than the measured line widths. Table \ref{tab:observations} reports the spectral line parameters derived for each emission line, in every observed position.

\begin{figure*}[ht]
\centering
\includegraphics[scale=0.50]{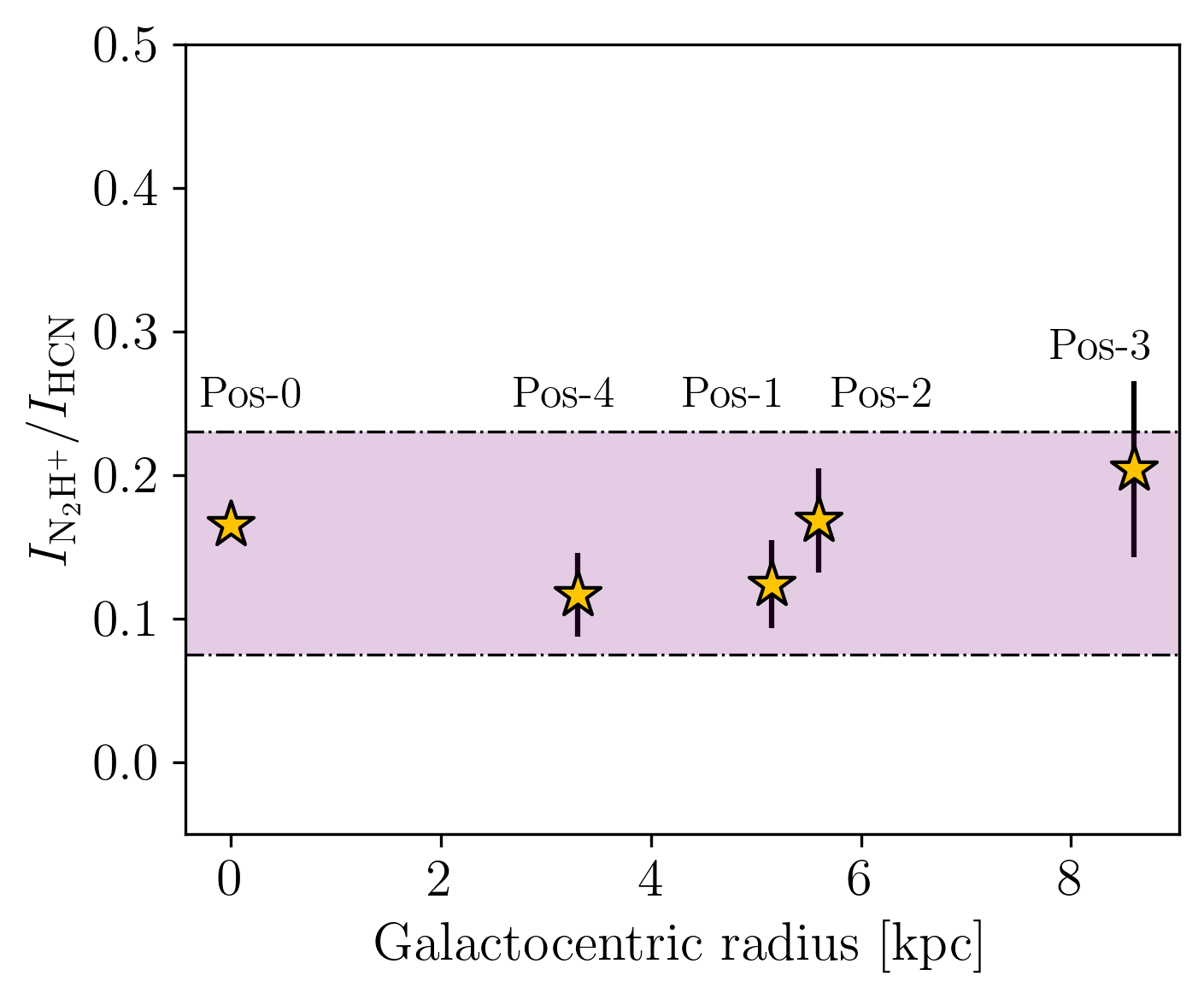}\,\includegraphics[scale=0.50]{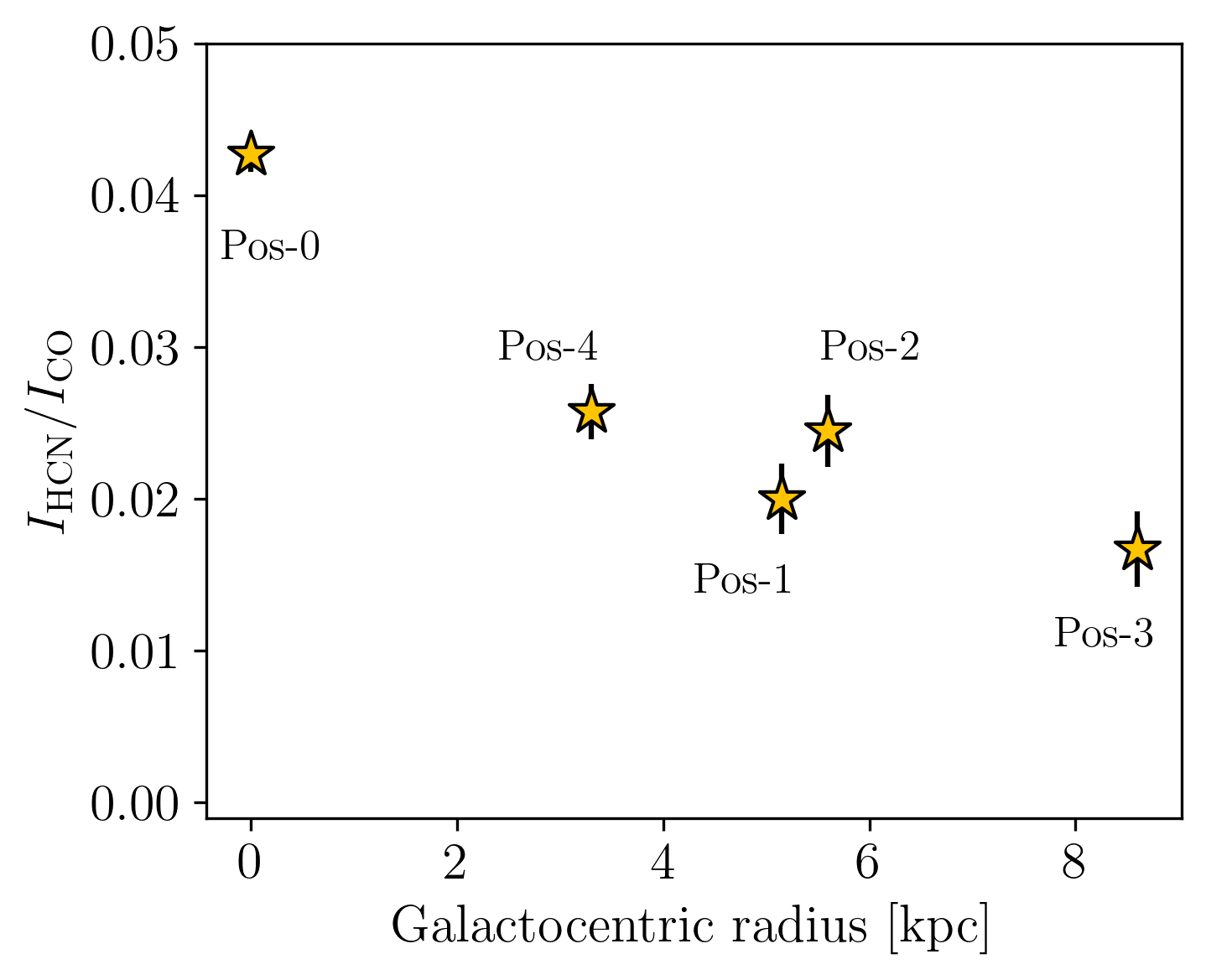}\,\includegraphics[scale=0.50]{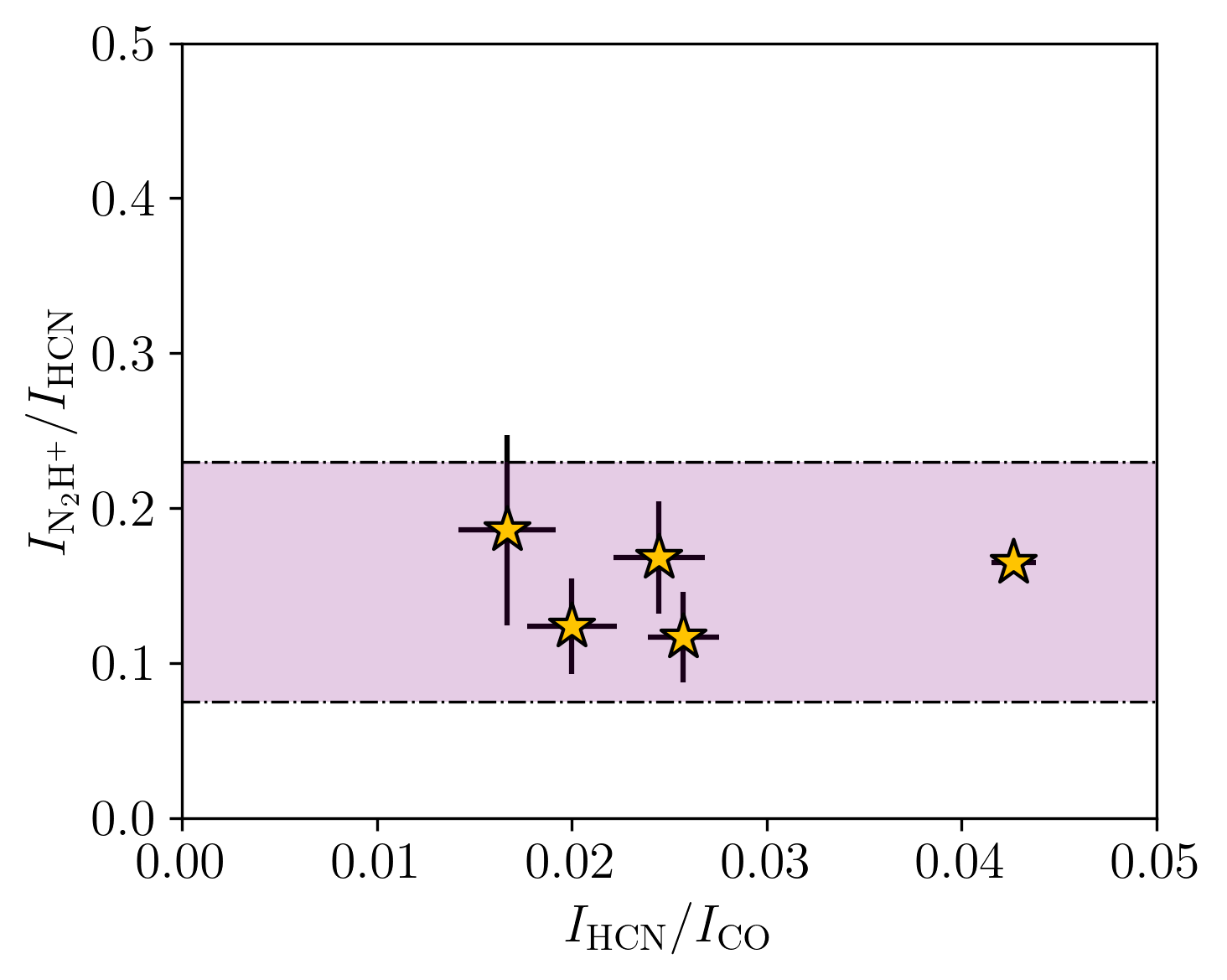}\\
\caption{Calculated $I_{\mathrm{N}_2\mathrm{H}^+}/I_{\mathrm{HCN}}$ ratios (left) from our measurements and $I_{\mathrm{HCN}}/I_{\mathrm{CO}}$ ratios (middle) from   EMPIRE survey data \citep{Donaire2019}, as a function of galactocentric radius in NGC\,6946. Each observed position is indicated. The right panel shows $I_{\mathrm{N}_2\mathrm{H}^+}/I_{\mathrm{HCN}}$ as a function of $I_{\mathrm{HCN}}/I_{\mathrm{CO}}$, illustrating that $I_{\mathrm{N}_2\mathrm{H}^+}/I_{\mathrm{HCN}}$ remains approximately constant, while $f_\mathrm{dense}$ changes by a factor of $\sim 4$. The purple region shows the range of previous literature values found for the central regions of starburst galaxies, AGNs, and ULIRGs \citep[e.g.,][but see also Fig. \ref{fig:literature}.]{Mauersberger1991,Watanabe2014,Aladro2015,Nishimura2016a,Eibensteiner2022}.}
\label{fig:observed_ratios}
\end{figure*}

\begin{figure*}[ht]
\centering
\includegraphics[scale=0.33]{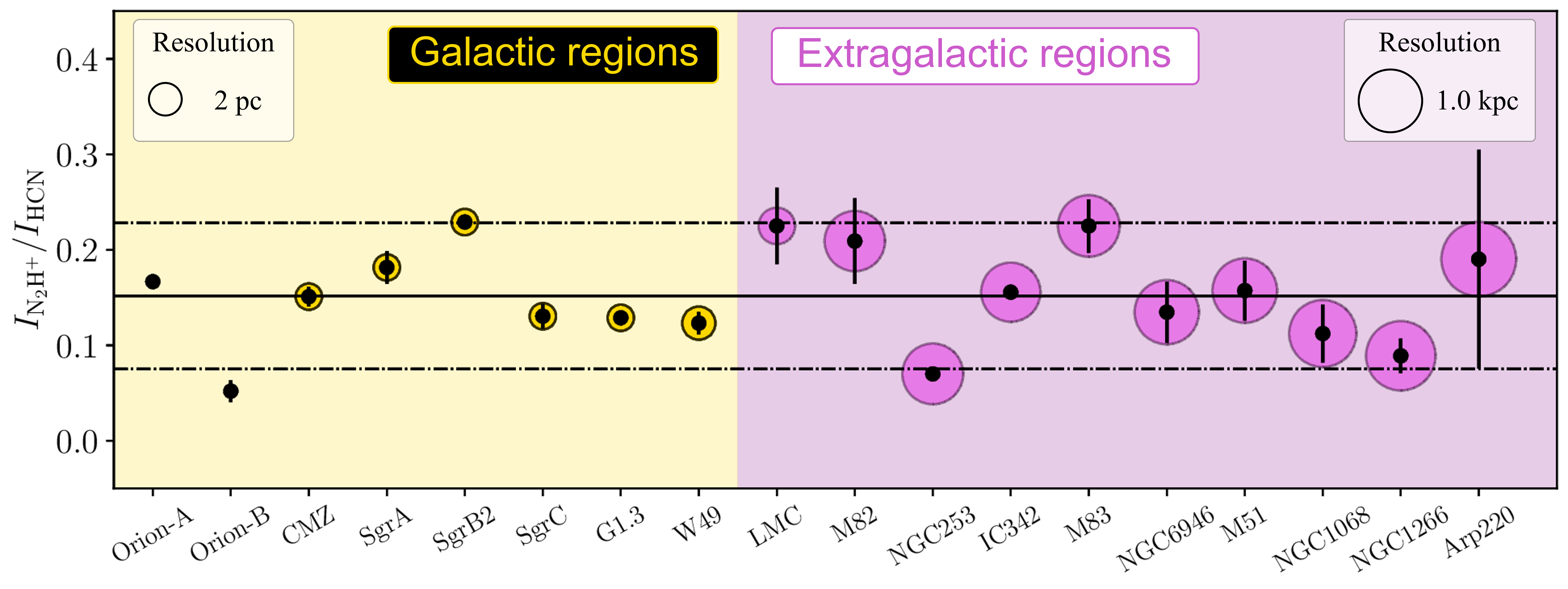}
\caption{$I_{\mathrm{N}_2\mathrm{H}^+}/I_{\mathrm{HCN}}$ ratios observed  toward nearby galaxies \citep[][and this work]{Aladro2015,Watanabe2014,Mauersberger1991,Nguyen1992,Eibensteiner2022,Nishimura2016a,Takano2019} and Galactic molecular clouds \citep{Jones2012,Pety2017,Barnes2020,Yun2021}. For galaxies with several measurements, we represent their mean values with associated uncertainties. The sizes of the circular markers are given in logarithmic scale. For visual reasons,  the markers for the Galactic regions are scaled with 500 and the markers in the extragalactic regions with 300. The solid line represents the mean value among all samples, while the dashed lines indicate a $3\sigma$ dispersion from the mean.}
\label{fig:literature}
\end{figure*}

The left panel of Fig.\,\ref{fig:observed_ratios} shows our estimated N$_2$H$^+$\,(1-0)-to-HCN\,(1-0) intensity ratios ($I_{\mathrm{N}_2\mathrm{H}^+}/I_{\mathrm{HCN}}$) as a function of galactocentric radius and observed position. We measure $I_{\mathrm{N}_2\mathrm{H}^+}/I_{\mathrm{HCN}}$ ratios that range between 0.12 and 0.20 in NGC\,6946, with a mean value of $0.15\pm0.02$ and a standard deviation of 0.03. Our line ratios agree well with the range of observed $I_{\mathrm{N}_2\mathrm{H}^+}/I_{\mathrm{HCN}}$ values in nearby starbursts, AGNs, and ULIRGs (see also Fig.\,\ref{fig:literature}), as reflected by the purple area in the plot. This compilation is available in Table\,\ref{tab:lit_data}. The middle panel of Fig.\,\ref{fig:observed_ratios} shows the HCN\,(1-0)-to-CO\,(1-0) intensity ratios ($I_{\mathrm{HCN}}/I_{\mathrm{CO}}$) from EMPIRE \citep{Donaire2019} extracted at the same positions.

\begin{figure*}[ht]
\includegraphics[scale=0.76]{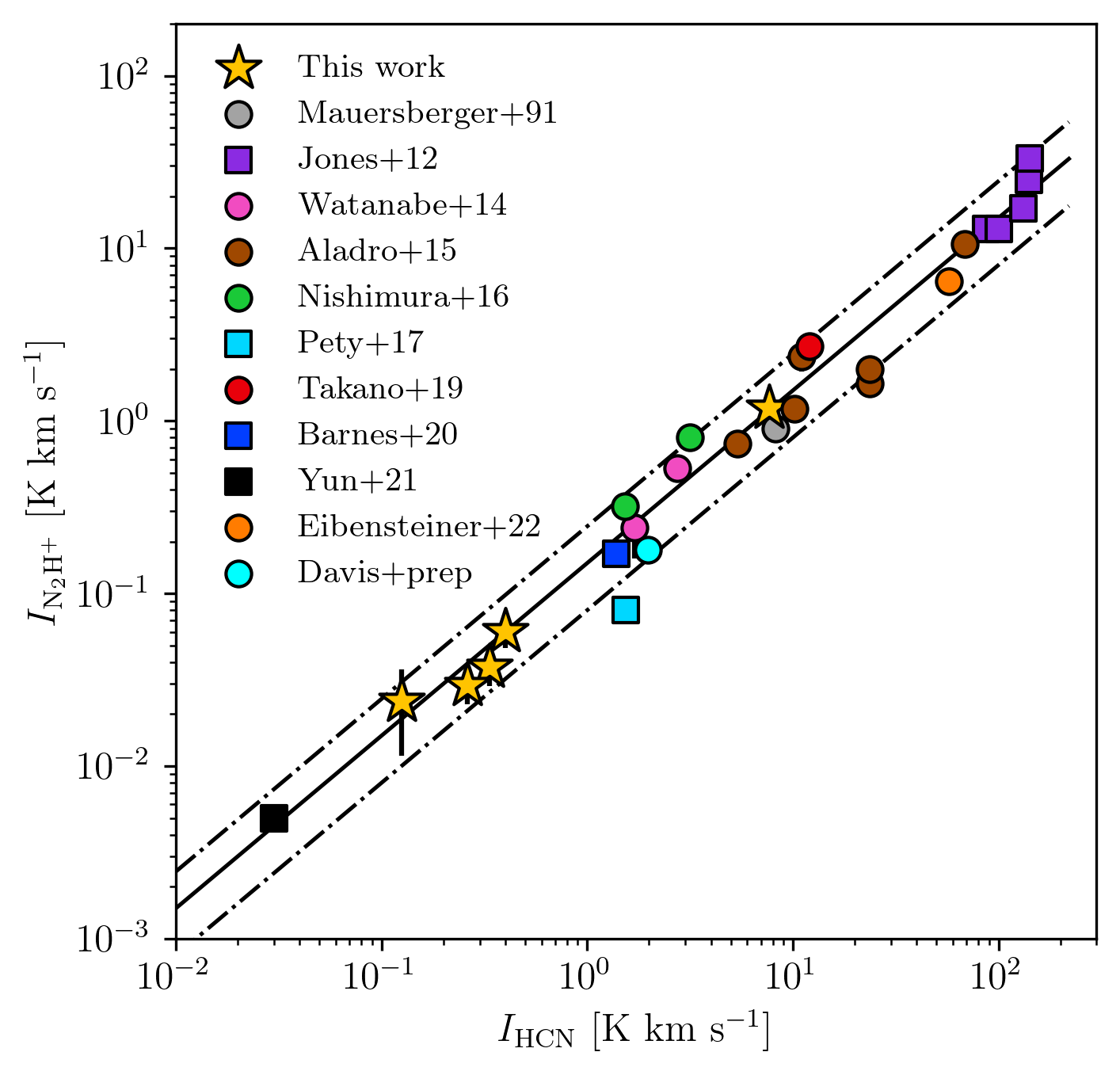}\,
\includegraphics[scale=0.76]{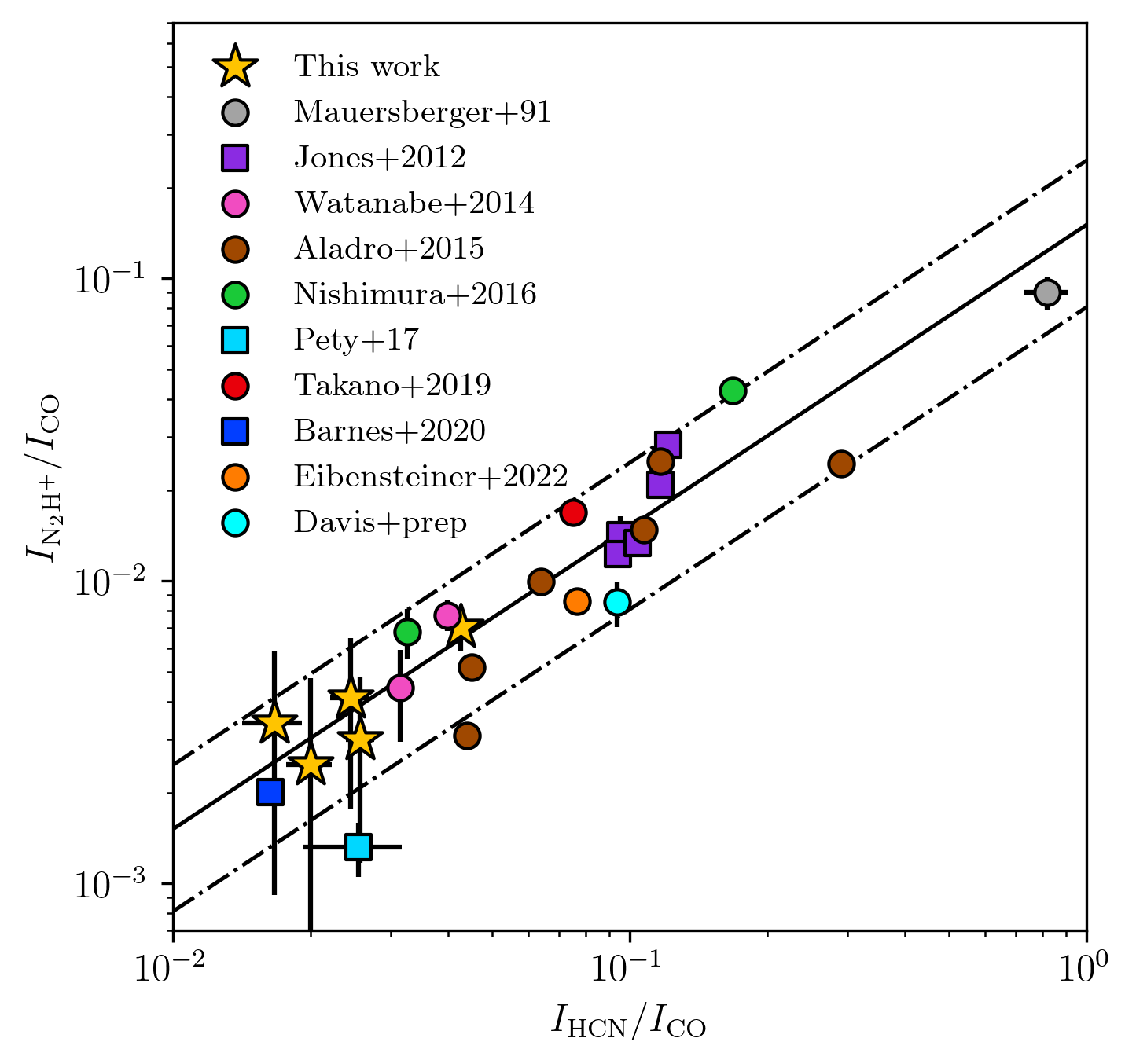}
\caption{Correlation between N$_2$H$^+$\,(1-0), HCN\,(1-0) and CO\,(1-0) intensities and line ratios. Left: Comparison between $I_{\mathrm{N}_2\mathrm{H}^+}$ and $I_{\mathrm{HCN}}$ in our observed locations (yellow stars), including available data for nearby galaxies \citep[circles,][]{Aladro2015,Watanabe2014,Mauersberger1991,Nguyen1992,Eibensteiner2022,Nishimura2016a,Takano2019} and Galactic molecular clouds \citep[squares,][]{Jones2012,Pety2017,Barnes2020,Yun2021}.  The solid line represents a linear fit to the ensemble data in both panels, showing that the two emission lines are strongly correlated ($r>0.98$) across almost four orders of magnitude. The dashed lines indicate a $3\sigma$ dispersion from the mean. Right: $I_{\mathrm{N}_2\mathrm{H}^+}/I_{\mathrm{CO}}$ as a function of $I_{\mathrm{HCN}}/I_{\mathrm{CO}}$ for NGC\,6946 and the same literature compilation. The two ratios, both proxies for the dense gas fraction, also remain well correlated across almost three orders of magnitude.
}
\label{fig:correlation}
\end{figure*}

While $I_{\mathrm{HCN}}/I_{\mathrm{CO}}$ systematically decreases by a factor of $\sim 4$ with increasing galactocentric distance, $I_{\mathrm{N}_2\mathrm{H}^+}/I_{\mathrm{HCN}}$ appears to be constant across the galaxy disk. We compare our observations (yellow stars) to a compilation of literature measurements in the left panel of Fig.\,\ref{fig:correlation}, showing integrated intensities of N$_2$H$^+$\,(1-0) as a function of HCN\,(1-0). This includes measurements of resolved Galactic clouds \citep[squares,][]{Jones2012,Pety2017,Barnes2020,Yun2021} and kiloparsec-scale regions in nearby galaxies \citep[circles,][]{Mauersberger1991,Watanabe2014,Aladro2015,Nishimura2016a,Takano2019,Eibensteiner2022}. While these data are taken at different resolutions (from 0.1\,pc to 10\,kpc), we note that such measurements are scarce and constitute our only source of comparison. It is clear from the plot that the strong correlation between $I_{\mathrm{N}_2\mathrm{H}^+}$ and $I_{\mathrm{HCN}}$ is observed across a wide variety of systems. A fit to the global literature data (including this work) yields
\begin{equation}
\mathrm{log}\,I_{\mathrm{N}_2\mathrm{H}^+,\mathrm{global}} = (0.99\pm0.04)\,\mathrm{log}\,I_{\mathrm{HCN,global}}-(0.87\pm0.04).   
\end{equation} 

We measured the strength and significance of the correlation between the derived intensities with the Spearman rank correlation coefficient, $r=0.96$, with a $p-\mathrm{value} < 0.001$. We note that this strong correlation between $I_{\mathrm{N}_2\mathrm{H}^+}$ and $I_{\mathrm{HCN}}$ holds, while $I_{\mathrm{HCN}}$ (and $I_{\mathrm{N}_2\mathrm{H}^+}$) varies across almost three orders of magnitude. 

The right panel of Fig.~\ref{fig:correlation} shows the correlation of $I_{\mathrm{N}_2\mathrm{H}^+}/I_{\mathrm{CO}}$ as a function of the commonly used $f_\mathrm{dense}$ tracer, $I_{\mathrm{HCN}}/I_{\mathrm{CO}}$, for the same literature compilation. These plots indicate that the two line ratios are correlated across several orders of magnitude, suggesting that they are equally good at tracing variations in $f_\mathrm{dense}$. A fit to all available data yields 
\begin{equation}
\mathrm{log}\,\frac{I_{\mathrm{N}_2\mathrm{H}^+}}{I_\mathrm{CO}}_\mathrm{global} = (1.00\pm0.05)\,\mathrm{log}\,\frac{I_{\mathrm{HCN}}}{I_{\mathrm{CO}}}_\mathrm{global}-(0.84\pm0.06),
\end{equation}

\noindent with $r=0.99$, and $p-\mathrm{value} < 0.001$. Figure~\ref{fig:literature} shows the observed line ratio ranges for our literature compilation in detail, targeting nearby galaxies, starbursts, AGNs and ULIRGs, as well as Milky Way regions. The solid line in the plot indicates a mean value of $0.15\pm0.03$ for the global $I_{\mathrm{N}_2\mathrm{H}^+}/I_{\mathrm{HCN}}$ ratio. The dashed lines represent a $3\sigma$ deviation from this mean. Our average value for NGC\,6946 is in good agreement with the literature values, within the uncertainties. In this broad comparison, the Orion B cloud is the only clear outlier from the correlation with a $I_{\mathrm{N}_2\mathrm{H}^+}/I_{\mathrm{HCN}}$ ratio that is three times smaller than the average value of 0.15 found from the literature compilation. In contrast, our sampled ratios in NGC\,6946 are very compatible with those found in low-metallicity environments such as the Large Magellanic Cloud (LMC), where we would expect important differences in the chemical composition of the clouds due to its deficiency in N-bearing molecules \citep{Dufour1982}. Lower HCN/HCO$^+$ ratios have been measured in the low-metallicity galaxies LMC and IC~10 \citep{Nishimura2016b,Nishimura2016a,Patra2022}, compared to the normal star-forming disks sampled by EMPIRE \citep{Donaire2019}.

\section{Implications for extragalactic dense gas observations}

Several studies have questioned whether the low-$J$ transitions of HCN and HCO$^+$, commonly used high critical density tracers, actually trace cold dense gas emission \citep[e.g.,][]{Pety2017,Kauffmann2017,Tafalla2021}. While these transitions have high critical densities, using this parameter as an indication of dense gas is an oversimplification. Among other effects, \citet{Shirley2015} showed that radiative trapping and excitation lower the effective critical density of these lines by one or two orders of magnitude \citep[see also][]{Donaire2017_a}. This causes a large fraction of the HCN emission to be subthermally excited \citep{Leroy2017,Jones2023,GarciaRodriguez2023}. Moreover, electron excitation can significantly contribute to its emissivity \citep{Goldsmith2017}. In fact, HCN\,(1-0) has a higher critical density than that of N$_2$H$^+$\,(1-0) ($4.7\times10^5$ vs. $6.1\times10^4\,\mathrm{cm}^{-3}$ at 10\,K, respectively), but the effective density is slightly higher for N$_2$H$^+$\,(1-0), which is significantly less opaque than HCN\,(1-0) \citep[][see Table 1]{Shirley2015}, even though both lines show  hfs. Some studies focusing on a small number of local clouds find extended regions with low-density gas ($50-100\,\mathrm{cm}^{-3}$) that contribute significantly to the global emission seen in HCN and HCO$^+$ \citep{Evans2020}. In particular, \citet{Pety2017} and \citet{Kauffmann2017} find that the vast majority of the HCN and HCO$^+$ luminosities come from regions with $A_V<8\,\mathrm{mag}$, while most dense star-forming cores are seen at $A_V\geq8\,\mathrm{mag}$. Both studies also agree that N$_2$H$^+$ is able to selectively trace gas at much higher densities in Orion~A and Orion~B, where at least 50\% of its emission comes from regions with $A_V\geq16\,\mathrm{mag}$. 

In addition, N$_2$H$^+$ primarily originates from cold gas ($<20\,\mathrm{K}$). In contrast, HCN is often found to originate from gas with moderate temperatures ($>35\,\mathrm{K}$) \citep[e.g.,][]{Pety2017,Barnes2020}. One of the reasons why the N$_2$H$^+$ molecule is a particularly good tracer of cold dense gas is that it is destroyed by CO via ion-neutral interactions \citep{Meier2005}. In addition, N$_2$H$^+$ formation depends on the available amount of $\mathrm{H}^+_3$ to react with $\mathrm{N}_2$, a process in which $\mathrm{N}_2$ competes with CO. In  cold high-density gas, CO freezes out onto the dust grains, eliminating the main N$_2$H$^+$ destroyer and making available more $\mathrm{H}^+_3$ to react with N$_2$ \citep{Caselli2012}. The combined result is a significant enhancement of the N$_2$H$^+$ abundance in cold and dense gas. 
Since the excitation conditions are similar for N$_2$H$^+$\,(1-0) and HCN\,(1-0), it is the depletion of CO that makes the N$_2$H$^+$\,(1-0) emission an ideal tracer of cold dense gas. Consequently, N$_2$H$^+$ yields a bright emission line in the densest regions of molecular clouds \citep{Caselli1999,Bergin2007}, while in those same regions carbon-bearing molecules freeze onto the dust grains \citep{Tafalla2002,Bergin2007}. Recent Galactic work has shown that while HCN\,(1-0) extends over a significant fraction of the cloud, it scales linearly with the H$_2$ column density of the gas \citep[e.g.,][]{Tafalla2021,Dame2023ApJ}. In NGC\,6946, cloud-to-cloud variations are averaged out at the scales we probe, yielding N$_2$H$^+$ proportional to HCN emission. If indeed N$_2$H$^+$ can selectively pick the densest and coldest regions (with a typical $A_V\geq16\,\mathrm{mag}$) and HCN mainly traces gas in regions with $A_V\sim 6-8\,\mathrm{mag}$, our remarkably constant $I_{\mathrm{N}_2\mathrm{H}^+}/I_{\mathrm{HCN}}$ ratio across the disk of NGC\,6946 suggests that cloud structures are similar and  subcloud differences average out at kiloparsec scales, yielding the two tracers proportional to each other.

The implications for extragalactic work are large. Over the last two decades, many studies have extended the strong correlation found by \citet{Gao2004} between HCN luminosity and IR-traced SFRs, from resolved galaxy disks \citep[e.g.,][]{Usero2015,Gallagher2018,Donaire2019} to individual cores in the Galaxy \citep[e.g.,][]{Wu2005,Wu2010,Stephens2016,Shimajiri2017}. Using HCN\,(1-0) to trace dense gas, \citet{Bigiel2016}, \citet{Donaire2019}, and \citet{Neumann2023}, among others, find systematic variations in $f_\mathrm{dense}$ and SFE$_\mathrm{dense}$ as a function of galactic environment. These results, however, are subject to the interpretation of the HCN emissivity. Our observations show a clear correlation between $I_{\mathrm{N}_2\mathrm{H}^+}$ and $I_{\mathrm{HCN}}$ within a prototypical star-forming galaxy, NGC\,6946, over a wide range of galactic radii, in regions with different morphologies (circumnuclear regions, spiral arms, and interarm region). These results are also in agreement with existing measurements for starbursts, ULIRGs, AGNs, and individual Galactic regions, where we find a constant ratio across more than three orders of magnitude in integrated intensities. This constant $I_{\mathrm{N}_2\mathrm{H}^+}/I_{\mathrm{HCN}}$ provides evidence that the variation of HCN\,(1-0) emission could be an interesting tool to estimate the cold dense gas traced by N$_2$H$^+$\,(1-0) across approximately kiloparsec-size regions. Moreover, ratios of other lines with different critical densities such as $I_{\mathrm{HCN}}/I_{\mathrm{CO}}$ and $I_{\mathrm{N}_2\mathrm{H}^+}/I_{\mathrm{CO}}$ can interchangeably be used as good indicators for $f_\mathrm{dense}$ in other galaxies, at least at kiloparsec scales, as previously pointed out by \citet{Leroy2017} and \citet{Gallagher2018b}. Additional kiloparsec-scale observations, however, are needed to confirm the observed trends in other nearby galaxies, including galaxy outskirts; high-resolution observations are also needed.

The strong correlation we observe implies that the amount of gas traced by HCN\,(1-0) is proportional to the amount of star-forming high-density gas traced by N$_2$H$^+$\,(1-0) at kiloparsec scales. This is expected if the structure of individual star-forming molecular clouds is  similar overall among different clouds, and is scale-free. The shape of the probability distribution of column densities ($N$-PDF) in the  observed molecular clouds is best described by power-law functions \citep{Kainulainen2009,Schneider2013,Lombardi2015,Abreu2015} in their high-density regimes. Kiloparsec-sized extragalactic observations capture ensembles of molecular clouds, where we expect the resulting PDF of the ensemble to reflect this power-law shape if the individual clouds are indeed self-similar (e.g., independent of the scale). In that case, the ratio of gas traced by HCN\,(1-0) and N$_2$H$^+$\,(1-0) (sensitive to moderate to high column density thresholds) is expected to be relatively constant. N$_2$H$^+$ is selective of cold dense gas, that is expected to ultimately form stars. It is interesting to find bright N$_2$H$^+$ emission in regions where HCN is also present, because HCN often traces moderately dense warmer gas (e.g., \citealp{Barnes2020}). This suggests that when observing a large region of another galaxy, it is common to find small highly shielded regions where conditions are conducive for N$_2$H$^+$ to be abundant, likely similar to the dark clouds found in our Galaxy. Our observations suggest that these regions have an approximately constant proportion to the warmer more excited dense gas, perhaps associated with photodissociation regions (PDRs), that likely dominate HCN emission. This could make sense given that the PDRs are natural products of star formation, and the conditions where N$_2$H$^+$ forms may be a relatively universal prerequisite to form stars. This adds significant weight to the interpretation of extragalactic HCN measurements in terms of tracing dense star-forming gas and the associated $f_\mathrm{dense}$ and $\mathrm{SFE}_\mathrm{dense}$.

\section{Summary and conclusions}

In this work we study the relation between N$_2$H$^+$\,(1-0) and HCN\,(1-0) across the disk of NGC\,6946, based on extensive observations with the IRAM-30m dish at $\sim90\,\mathrm{GHz}$. We find that the integrated intensities from the two emission lines are strongly correlated within the galaxy disk at kiloparsec scales. Moreover, this correlation is also extended to a wide range of physical environments including Galactic clouds, nearby starburst galaxies, AGNs, ULIRGs, and even low-metallicity regimes. Dense gas fractions, as probed by $I_{\mathrm{HCN}}/I_{\mathrm{CO}}$ and $I_{\mathrm{N}_2\mathrm{H}^+}/I_{\mathrm{CO}}$, are also very well correlated across several orders of magnitude. Our results directly address recent discussions of varying $f_\mathrm{dense}$ and $\mathrm{SFE}_\mathrm{dense}$ within and across galaxies, and indicate that despite the known caveats for HCN emission, this molecule can be used instead of canonical Galactic dense gas tracers like N$_2$H$^+$ to estimate $f_\mathrm{dense}$, at least on kiloparsec scales.

\begin{acknowledgements}
      This work is based on observations carried out with the IRAM-30m telescope. IRAM is supported by INSU/CNRS (France), MPG (Germany) and IGN (Spain). The authors would like to E.\,Pellegrini, D.\,Harsono and S.\,Ellison for useful discussions, and the referee, Neal Evans, for a constructive and helpful report. AH acknowledges funding from the European Research Council (ERC) under the European Union’s Horizon 2020 research and innovation programme (Grant agreement Nos. 851435). RSK and SCOG acknowledge financial support from the ERC via the ERC Synergy Grant ``ECOGAL'' (project ID 855130), from the German Excellence Strategy via the Heidelberg Cluster of Excellence (EXC 2181 - 390900948) ``STRUCTURES'', and from the German Ministry for Economic Affairs and Climate Action in project ``MAINN'' (funding ID 50OO2206). RSK also thanks for computing resources provided bwHPC and DFG through grant INST 35/1134-1 FUGG and for data storage at SDS@hd through grant INST 35/1314-1 FUGG. MJJD and MQ acknowledge support from the Spanish grant PID2019-106027GA-C44, funded by MCIN/AEI/10.13039/501100011033. MC gratefully acknowledges funding from the DFG through an Emmy Noether Research Group (grant number CH2137/1-1). LN acknowledges funding from the Deutsche Forschungsgemeinschaft (DFG, German Research Foundation) - 516405419. AU acknowledges support from the Spanish grants PGC2018-094671-B-I00, funded by MCIN/AEI/10.13039/501100011033 and by ``ERDF A way of making Europe'', and PID2019-108765GB-I00, funded by MCIN/AEI/10.13039/501100011033. 
\end{acknowledgements}

\bibliographystyle{aa}
\bibliography{biblio}

\begin{appendix}

\section{Observations and data analysis}
\label{ap:datareduction}

The data were taken as part of projects 061-18, 055-20, 054-20, and 159-20 (PI: M. J. Jim\'enez-Donaire). The E090 receiver yielded an instantaneous bandwidth of 15.6\,GHz per polarization, and used the fast Fourier transform (FFT) spectrometers with 195\,kHz spectral resolution (FTS200, $\sim 0.5\,\mathrm{km}\,\mathrm{s}^{-1}$) to record the data. The observations were performed in five different positions, representative of distinct environments within the nearby galaxy NGC\,6946 (see Fig.\,\ref{fig:composite}). These were carried out in wobbler-switching mode with a total ON-OFF throw of $\pm110\arcsec$ in azimuth and a wobbling frequency of 0.5\,Hz. Four   of these positions were previously observed by \citet{Usero2015}, but those observations were not sensitive enough to detect N$_2$H$^+$\,(1-0). Every position was observed only when the line emission in the OFF positions along the Azimuth axis could not produce significant contamination. We did this by checking the integrated intensity and velocity in the HERACLES CO\,(2-1) cube of the target galaxy \citep{Leroy2009}. We tuned EMIR with a local oscillator frequency of $\sim 88.07\,\mathrm{GHz}$, which allowed us to simultaneously observe HCN\,(1-0) in the lower-outer subband, and N$_2$H$^+$\,(1-0) transition in the lower-inner subband. Additionally, other fainter transitions, such as dense gas isotopologs (H$^{13}$CN\,(1-0), H$^{13}$CO$^+$\,(1-0), and HN$^{13}$C\,(1-0)), are present in the observed band and will be analyzed in a future paper. The angular resolution of the IRAM-30m at the frequency of HCN\,(1-0) is $27.9\arcsec$. This sets our working resolution, which corresponds to a spatial resolution of $\sim 0.92\,\mathrm{kpc}$ at the distance of NGC\,6946. Because the observations of N$_2$H$^+$\,(1-0) and HCN\,(1-0) have similar beam sizes, aperture corrections are not applied.

During the course of the observations, the focus of the telescope was checked on planets (Mars and Saturn) or bright quasars (K3-50A) at the beginning of each session and then every $\sim 3$\,hours. In addition, it was also checked at sunset and sunrise if needed. We corrected the telescope pointing approximately every 1--1.5 hours using point-like sources close to NGC\,6946, such as 2037+511, 1226+023, 1928+738, and 1253-055. We obtained a standard chopper wheel calibration every $\sim 15\,$minutes to place the data on the antenna temperature scale ($T_\mathrm{A}^*$). The weather conditions were good overall, and the mean radiometer opacity at 225\,GHz was measured at around 0.5.

\begin{figure*}[ht]
\includegraphics[scale=0.55]{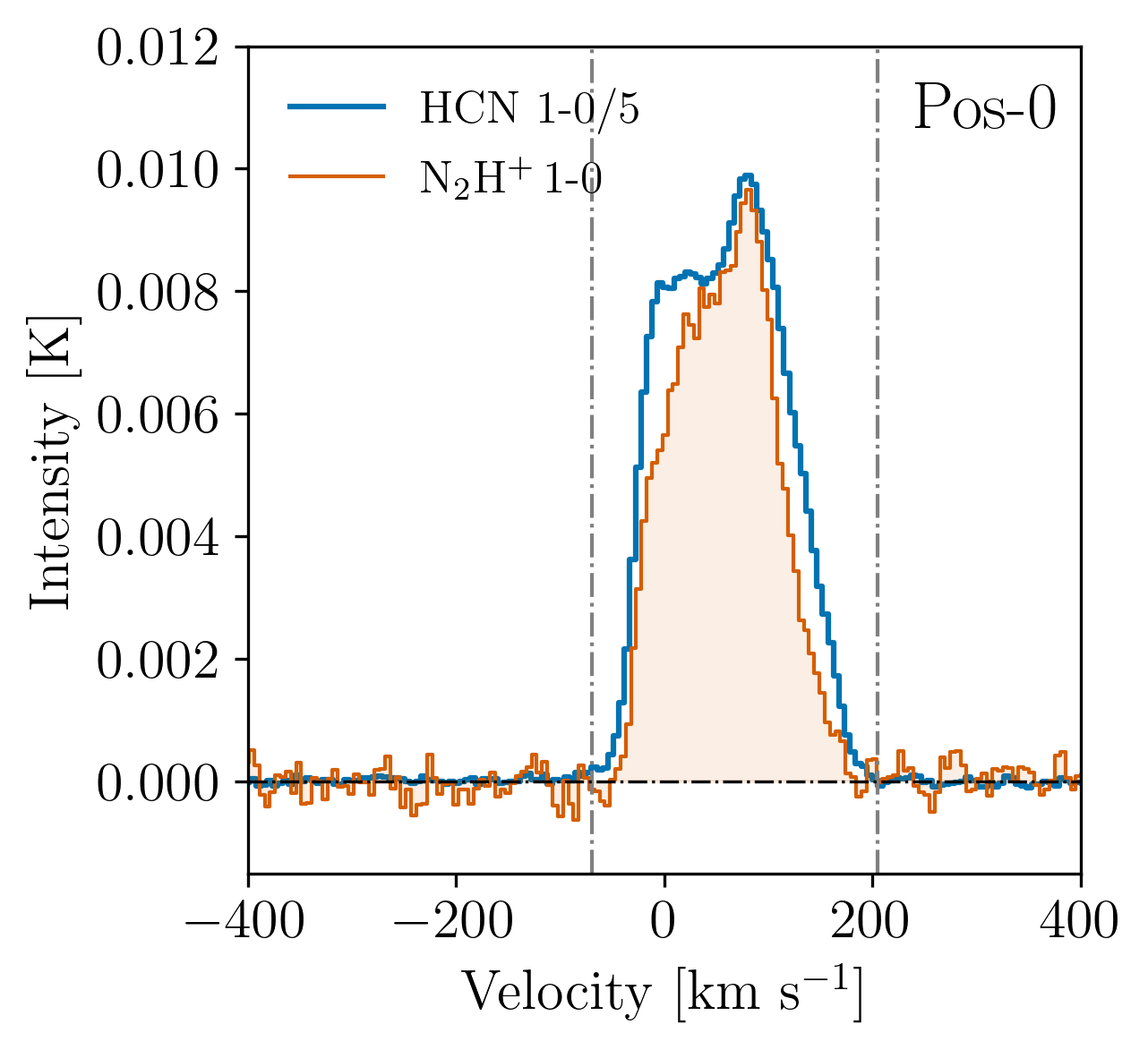}\,
\includegraphics[scale=0.55]{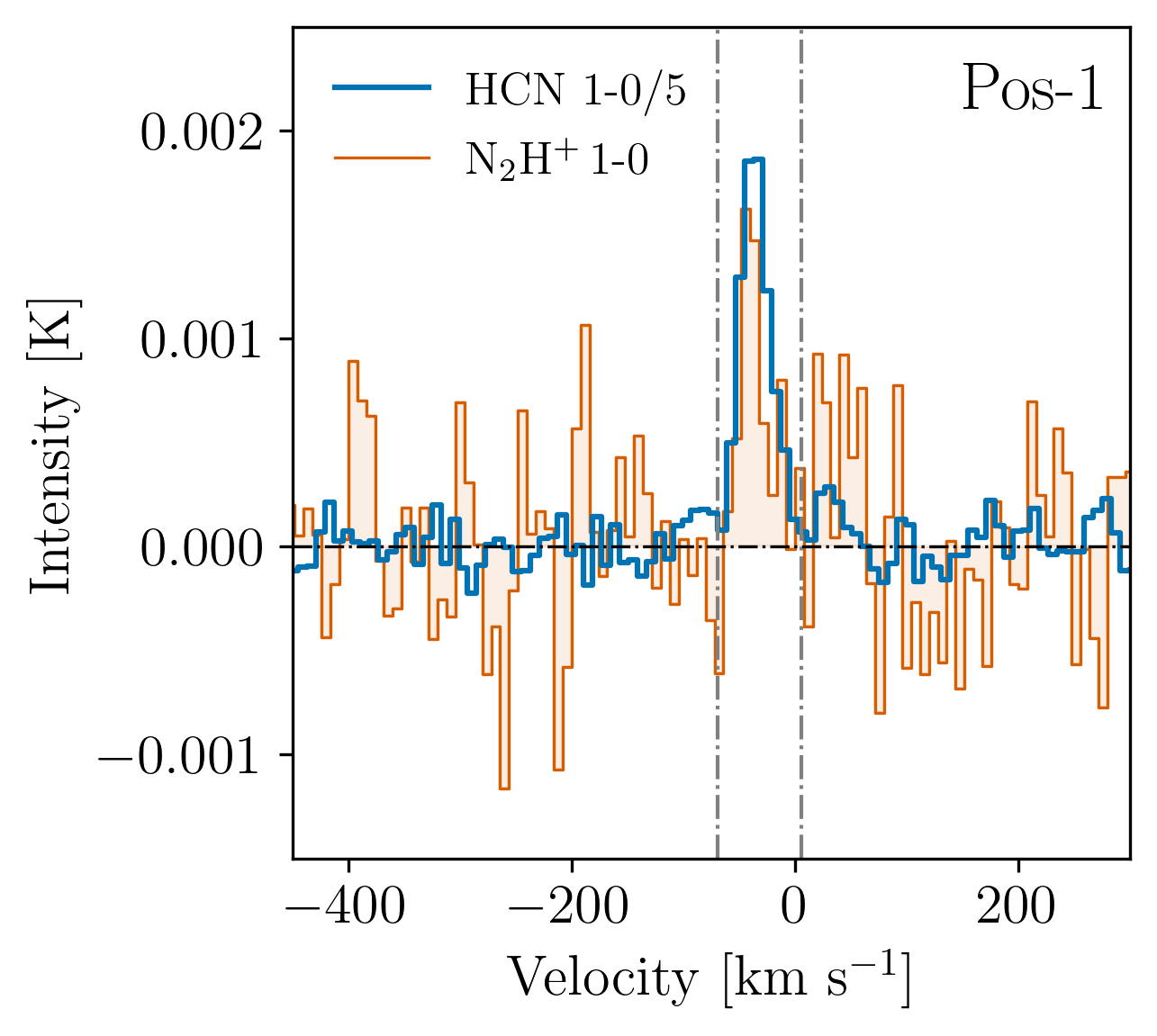}\,
\includegraphics[scale=0.55]{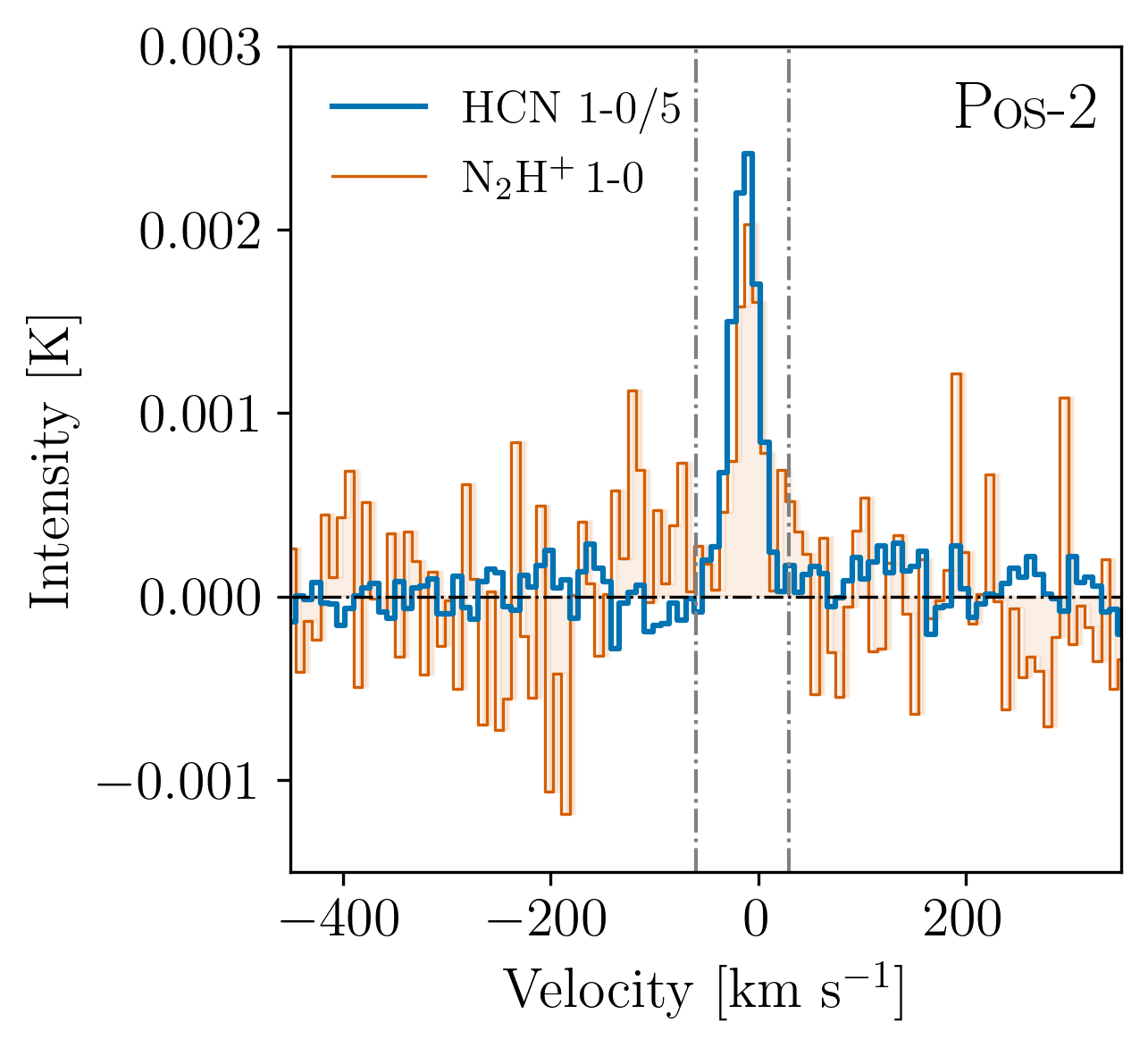}\\
\includegraphics[scale=0.55]{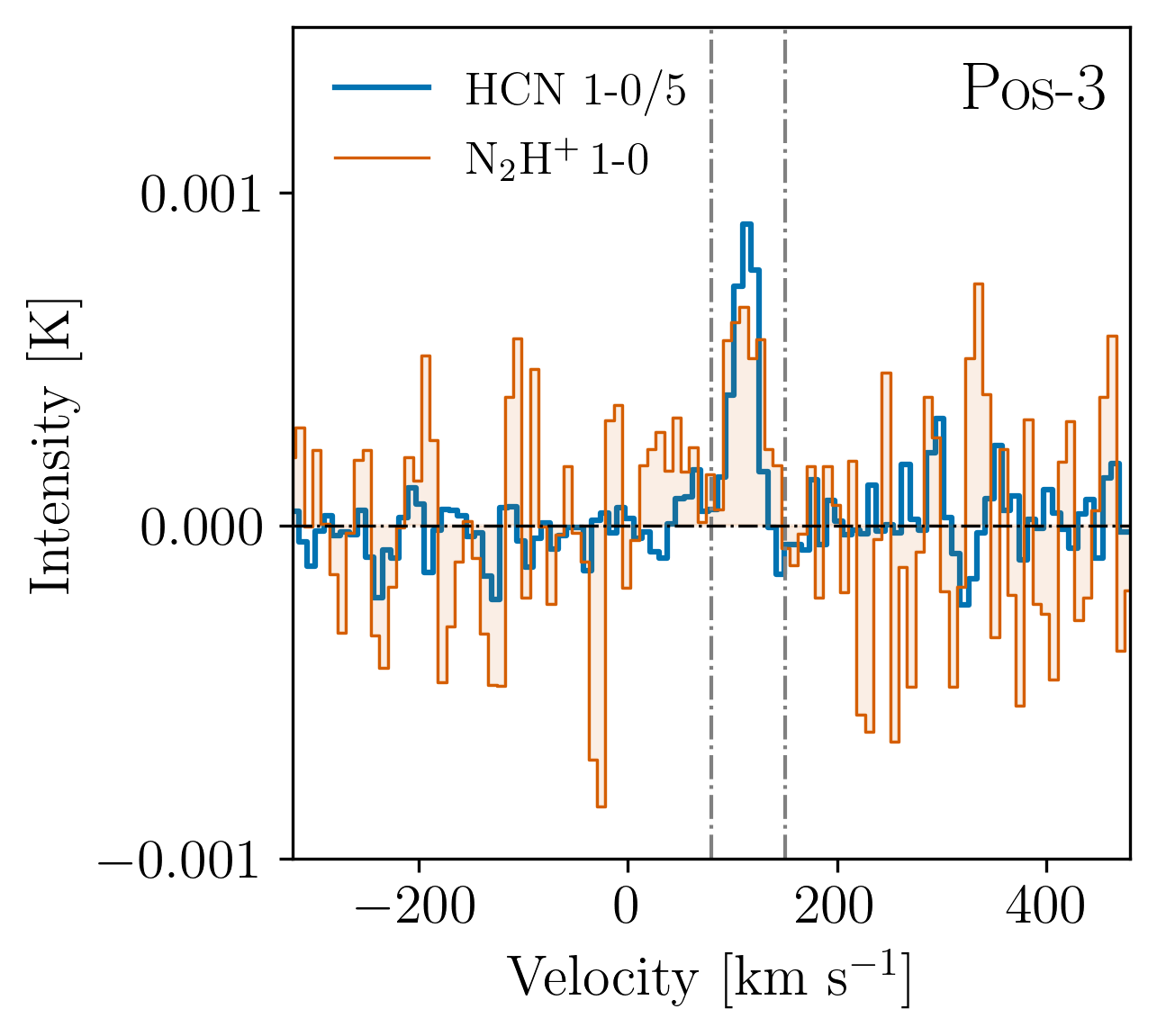}\,
\includegraphics[scale=0.55]{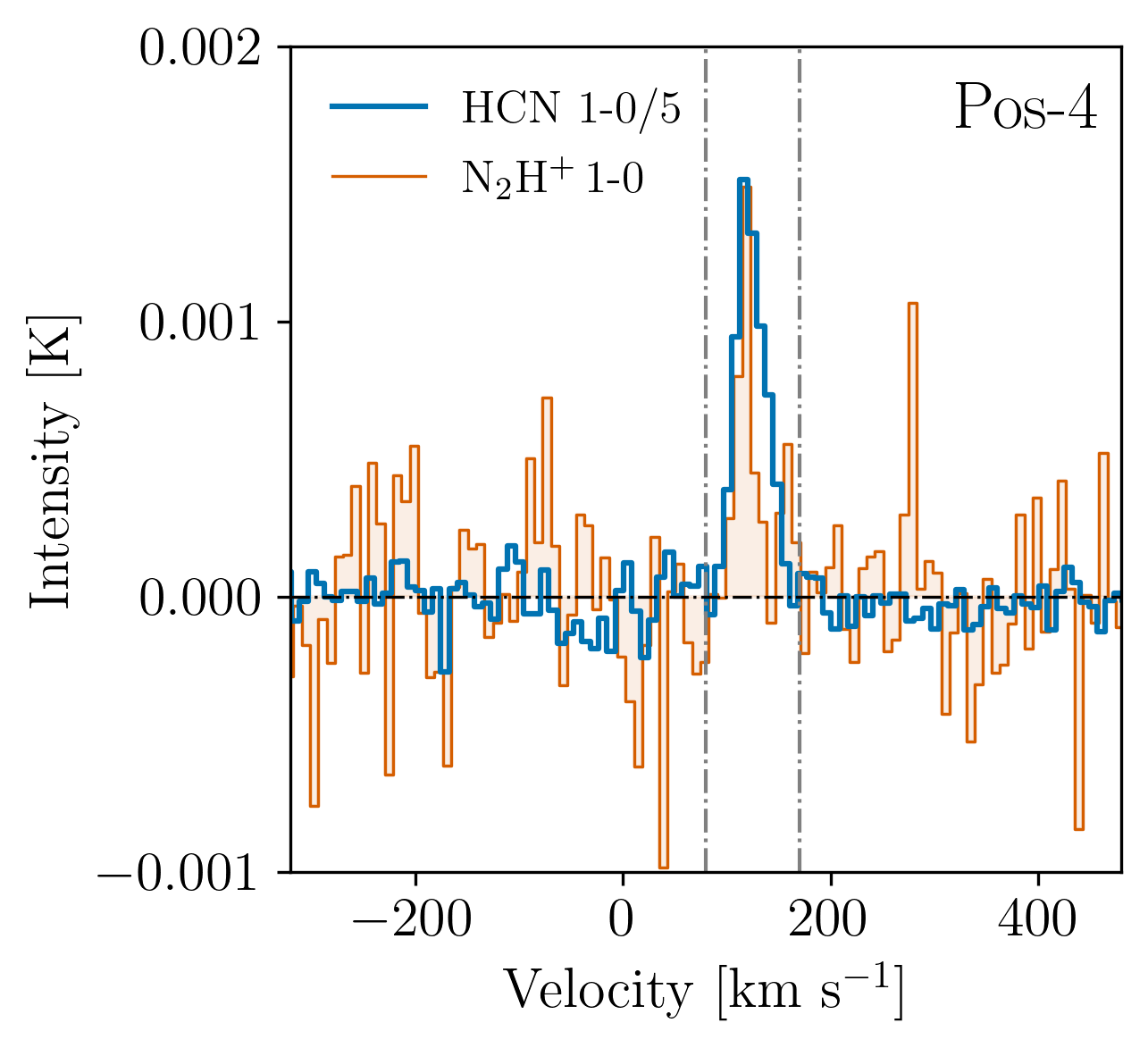}\
\caption{Individual spectra for each observed position as a function of the LSR velocity. The orange lines show the N$_2$H$^+$\,(1-0) spectrum, while the blue lines show the HCN\,(1-0) emission divided by a factor of 5. The vertical dashed lines delimit the windows used to reduce the data and derive the spectral line parameters. N$_2$H$^+$\,(1-0) shows good agreement with the mean HCN LSR velocity. Table \ref{tab:observations} reports the derived spectral line parameters.}
\label{fig:spectra}
\end{figure*}

The spectra for each observed location were calibrated with {\tt MRTCAL}\footnote{\url{https://www.iram.fr/IRAMFR/GILDAS/doc/pdf/mrtcal-prog-manual.pdf}} as part of
the {\tt GILDAS}\footnote{\url{https://www.iram.fr/IRAMFR/GILDAS/}} software package. The data were then converted to main beam temperature ($T_\mathrm{MB}$) with the available beam efficiencies from the IRAM documentation\footnote{\url{https://publicwiki.iram.es/Iram30mEfficiencies}} ($F_{\mathrm{eff}}=95\%$ and $B_{\mathrm{eff}}=81\%$ at 3 mm) following equation $T_\mathrm{MB} = F_{\mathrm{eff}}/B_{\mathrm{eff}} \times T_\mathrm{A}^*$. We extracted the target emission lines using the Continuum and Line Analysis Single-dish Software\footnote{\url{https://www.iram.fr/IRAMFR/GILDAS/doc/html/class-html/class.html}} ({\tt CLASS}). We subtracted a third-order polynomial baseline in each spectrum, and avoided fitting the baseline within the velocity range of molecular emission by using the available CO velocity field maps from the EMPIRE survey. Thus, we omitted the range of $\pm150\,\mathrm{km}\,\mathrm{s}^{-1}$ centered around the galactic mean CO\,(1-0) velocity. The individual spectra were then re-gridded to have a $4\,\mathrm{km}\,\mathrm{s}^{-1}$ channel width across the full bandpass, and averaged to produce a final spectrum per observed location. For each position, we calculated the integrated intensities by summing over the velocity range containing bright molecular emission HCN\,(1-0) (see dashed limits in Fig.\,\ref{fig:spectra}). Finally, we measured the root mean square (rms) noise level of $T_\mathrm{MB}$ at line-free channels of the HCN and N$_2$H$^+$ spectra. These values and detailed information about the observations can be found in Table \ref{tab:observations}.

Figure\,\ref{fig:spectra} presents the resulting individual spectra obtained for the five different positions shown in Fig.\,\ref{fig:composite}, plotted in main-beam brightness temperatures and smoothed to a common spectral resolution of $4\,\mathrm{km}\,\mathrm{s}^{-1}$. The average rms noise level achieved by the observations is 0.46\,mK for the HCN\,(1-0) spectra and 0.34\,mK for N$_2$H$^+$\,(1-0).

\section{Literature data}
\label{ap:lit_data}

The data presented in Table\,\ref{tab:lit_data} contains the most up-to-date N$_2$H$^+$\,(1-0) and HCN\,(1-0) observations in the literature. This compilation includes the data used to construct Fig.\,\ref{fig:correlation} and Fig.\ref{fig:literature}. Upper limits given by \citet{Nishimura2016b} were not employed in the plots. When this compilation table is used, the original studies providing the various data sets should be referenced.

\begin{table}
    \begin{center}
    \caption{{Galactic and extragalactic N$_2$H$^+$\,(1-0) and HCN\,(1-0) line measurements.}}
    \label{tab:lit_data}
    \begin{tabular}{l r r r}
    \hline \hline
    Source              & $I_{\mathrm{N}_2\mathrm{H}^+}$ & $I_{\mathrm{HCN}}$ & Resol.\\
    & [K~km~s$^{-1}$] & [K~km~s$^{-1}$] & \\ \hline
    Orion-A$^{(1)}$     & 0.005 & 0.03 & 0.11\,pc \\
    Orion-B$^{(2)}$     & $0.08\pm0.01$ & $1.5\pm0.2$ & 0.05\,pc \\
    CMZ$^{(3)}$         & $32\pm2$ & $212\pm1$ & 2\,pc \\
    SgrA$^{(3)}$        & $5.7\pm0.5$ & $31.4\pm0.2$ & 2\,pc \\
    SgrB2$^{(3)}$       & $9.3\pm0.2$ & $40.6\pm0.3$ & 2\,pc \\
    SgrC$^{(3)}$        & $2.4\pm0.2$ & $18.4\pm0.3$ & 2\,pc \\
    G1.3$^{(3)}$        & $17.0\pm0.2$ & $132.0\pm0.3$ & 2\,pc \\
    W49$^{(4)}$         & $0.17\pm0.01$ & $1.38\pm0.06$ & 3\,pc \\
    IC\,10$^{(5)}$     & $<0.11$ & $0.34\pm0.01$ & 4.9\,pc \\
    LMC-N113$^{(6)}$    & $0.80\pm0.07$ & $3.15\pm0.05$ & 9.2\,pc \\
    LMC-N159W$^{(6)}$   & $0.32\pm0.06$ & $1.53\pm0.04$ & 9.2\,pc \\
    IC342$^{(7)}$       & $2.1\pm0.4$ & $26.4\pm0.6$ & 340\,pc \\
    M82$^{(8)}$         & $1.65\pm0.04$ & $23.6\pm0.4$ & 428\,pc \\
    NGC\,253$^{(8)}$    & $10.6\pm0.2$ & $68.4\pm0.7$ & 466\,pc \\
    M83$^{(8)}$         & $1.17\pm0.05$ & $10.2\pm0.1$ & 554\,pc \\
    NGC\,6946$^{(9,10)}$ & $0.90\pm0.02$ & $8.2\pm0.1$ & 957\,pc \\
    NGC\,6946$^{(11)}$  & $6.4\pm0.5$ & $57.4\pm0.6$ & 100\,pc \\
    M51$^{(8)}$         & $0.74\pm0.04$ & $5.4\pm0.1$ & 1\,kpc \\
    M51-P1$^{(12)}$     & $0.53\pm0.06$ & $2.7\pm0.1$ & 1\,kpc \\
    M51-P2$^{(12)}$     & $0.24\pm0.08$ & $1.7\pm0.1$ & 1\,kpc \\
    NGC\,1068$^{(8)}$   & $1.99\pm0.08$ & $23.55\pm0.28$ & 1.8\,kpc \\
    NGC\,1266$^{(13)}$  & $0.17\pm0.03$ & $1.89\pm0.04$ & 3\,kpc \\
    Arp220$^{(8)}$      & $2.36\pm0.42$ & $11.06\pm0.24$ & 10.5\,kpc \\
    \hline
    \hline \hline
    \end{tabular}
    \end{center}
    \begin{minipage}{0.95\columnwidth}
        \vspace{1mm}
        {\bf Notes:} (1): \citet{Yun2021} do not provide uncertainties; (2) \cite{Pety2017}; (3): \cite{Jones2012}; (4): \cite{Barnes2020}; (5): \cite{Nishimura2016b}; (6): \cite{Nishimura2016a}; (7): \cite{Takano2019}; (8): \citet{Aladro2015}; (9): \citet{Mauersberger1991}; (10): \citet{Nguyen1992}; (11): \citet{Eibensteiner2022}; (12): \citet{Watanabe2014}; (13): Davis et al. in prep.
    \end{minipage}
\end{table}

\end{appendix}

\end{document}